\documentclass[fleqn,twoside]{article}
\usepackage{amsfonts}

\newcommand{\bls}[1]{\renewcommand{\baselinestretch}{#1}}

\def\noi{\noindent}

\makeatletter
\renewcommand{\section}{\@startsection{section}{1}{0pt}%
        {-3.5ex plus -1ex minus -.2ex}{2.3ex plus .2ex}%
        {\large\bf\protect\raggedright}}

\renewcommand{\subsection}{\@startsection{subsection}{2}{0pt}%
        {-3ex plus -1ex minus -.2ex}{1.4ex plus .2ex}%
        {\normalsize\bf\protect\raggedright}}

\renewcommand{\@oddhead}{\raisebox{0pt}[\headheight][0pt]{%
   \vbox{\hbox to\textwidth{\rightmark \hfil \rm \thepage \strut}\hrule}}}
\renewcommand{\@evenhead}{\raisebox{0pt}[\headheight][0pt]{%
   \vbox{\hbox to\textwidth{\thepage \hfil \leftmark \strut}\hrule}}}

\newcommand{\Acknow}[1]{\subsection*{Acknowledgement} #1}

\makeatother

\newcommand{\Title}[1]{\noi {\Large #1} \\}

\newcommand{\Abstract}[1]{\vskip 2mm \begin{center}
        \parbox{16.4cm}{\small\noi #1} \end{center}\medskip}

\newcommand{\foom}[1]{\protect\footnotemark[#1]}
\newcommand{\foox}[2]{\footnotetext[#1]{#2}}
\newcommand{\email}[2]{\footnotetext[#1]{e-mail: #2}}

\newcommand{\Ref}[1]{Ref.\,\cite{#1}}
\newcommand{\sect}[1]{Sec.\,#1}
\def\nq{\hspace*{-1em}}
\def\nqq{\hspace*{-2em}}
\def\nhq{\hspace*{-0.5em}}

\def\cm{\hspace*{1cm}}

\def\Jl#1#2{{\it #1\/} {\bf #2},\ }

\def\CQG#1 {\Jl{Clas. Qu. Grav.}{#1}}
\def\DAN#1 {\Jl{Dokl. AN SSSR}{#1}}
\def\GC#1 {\Jl{Grav. \& Cosmol.}{#1}}
\def\GRG#1 {\Jl{Gen. Rel. Grav.}{#1}}
\def\JETF#1 {\Jl{Zh. Eksp. Teor. Fiz.}{#1}}
\def\JMP#1 {\Jl{J. Math. Phys.}{#1}}
\def\NP#1 {\Jl{Nucl. Phys.}{#1}}
\def\PLA#1 {\Jl{Phys. Lett.}{#1A}}
\def\PLB#1 {\Jl{Phys. Lett.}{#1B}}
\def\PRD#1 {\Jl{Phys. Rev.}{D\ #1}}
\def\PRL#1 {\Jl{Phys. Rev. Lett.}{#1}}

\def\al{&\nhq}
\def\lal{&&\nqq {}}
\def\eq{Eq.\,}

\def\beq{\begin{equation}}
\def\eeq{\end{equation}}
\def\bear{\begin{eqnarray}}
\def\bearr{\begin{eqnarray} \lal}
\def\ear{\end{eqnarray}}
\def\earn{\nonumber \end{eqnarray}}
\def\nn{\nonumber\\ {}}

\def\nnn{\nonumber\\ \lal }

\def\yy{\\[5pt] {}}

\def\eql{\al =\al}

\def\tst{\textstyle}

\def\fract#1#2{{\tst\frac{#1}{#2}}}

\def\eqdef{\stackrel{\rm def}=}
\def\e{{\,\rm e}}
\def\d{\partial}

\def\sign{\mathop{\rm sign}\nolimits}

\def\const{{\rm const}}

\def\DAL{\mathop{\raisebox{3.5pt}{\large\fbox{}}}\nolimits}
\newcommand{\vars}[1]{\left\{\begin{array}{ll}#1\end{array}\right.}

\def\GR{general relativity}
\def\sph{spherically symmetric}
\def\ssph{static, spherically symmetric}

\def\wh{wormhole}
\def\whs{wormholes}

\def\mn{_{\mu\nu}}
\def\MN{^{\mu\nu}}

\def\og{{\overline g}}

\def\M{{\mathbb M}}
\def\R{{\mathbb R}}
\def\S{{\mathbb S}}
\def\ME{\mbox{$\M_{\rm E}$}}
\def\MJ{\mbox{$\M_{\rm J}$}}

\def\umx{u_{\max}}
\def\phio{1/\sqrt{\xi}}
\def\Str{\mbox{$\S_{\rm trans}$}}

\bls{0.97}
\begin{document}
\twocolumn[
\thispagestyle{empty}
\rightline {\bf gr-qc/0205131}
\bigskip

\Title   {\uppercase{Charged wormholes with non-minimally coupled \yy
        scalar fields. Existence and stability\foom 1}}

\noi {\large\bf K. Bronnikov,$^{a,b,2}$ and S. Grinyok,$^{a,3}$}

\medskip
{\protect
\begin{description}\itemsep -1pt
\item[$^a$]{\it Institute of Gravitation and Cosmology, PFUR,
        6 Miklukho-Maklaya St., Moscow 117198, Russia}
\item[$^b$]{\it VNIIMS, 3-1 M.Ulyanovoy St., Moscow 117313, Russia}
 \end{description}}

\Abstract
{Static, spherically symmetric, traversable wormhole solutions with electric
or magnetic charges are shown to exist in general relativity in the presence
of scalar fields nonminimally coupled to gravity. These \whs, however, turn
out to be unstable under \sph\ perturbations. The instability is related to
blowing-up of the effective gravitational constant on a certain sphere.  }

]  
\foox 1 {Contribution to Festschrift in honour of Prof. Mario Novello}
\email 2 {kb@rgs.mccme.ru}
\email 3 {stepan@rgs.phys.msu.su}

\section{Introduction}

   A search for traversable \wh\ solutions to the gravitational field
   equations with realistic matter has been for long, and is still remaining
   to be, one of the most intriguing challenges in gravitational studies.
   One of attractive features of \whs\ is their ability to support electric
   or magnetic ``charge without charge'' \cite{wheeler} by letting the lines
   of force thread from one spatial asymptotic to another.

   As is widely known, traversable wormholes can only exist with exotic
   matter sources, more precisely, if the energy-momentum tensor (EMT) of
   the matter source of gravity violates the local and averaged null energy
   condition (NEC) $T\mn k^{\mu}k^{\nu} \geq 0$, $k_\mu k^\mu =0$
   \cite{hoh-vis}. It is known, for instance, that nonlinear electrodynamics
   with any Lagrangian of the form $L({\cal F})$, ${\cal F} = F\MN F\mn$,
   coupled to \GR\ cannot produce a \ssph\ \wh\ metric \cite{br01-ned}.
   Though, an effective \wh\ geometry for electromagnetic wave propagation
   can appear as a result of the electromagnetic field nonlinearity
   \cite{nov1, nov2}.

   Scalar fields are able to provide good examples matter needed for \whs: on
   the one hand, in many particular models they do exhibit exotic
   properties, on the other, many exact solutions are known for gravity with
   scalar sources. We will consider some examples of charged \wh\ solutions
   in the presence of massless scalar fields.

   Let us begin with the action for a general (Bergmann-Wagoner) class of
   scalar-tensor theories (STT), where gravity is characterized by the
   metric $g\mn$ and the scalar field $\phi$ in the presence of the
   electromagnetic field $F\mn$ as the only matter source:
\bearr            \nq\
    S = \int d^4 x \sqrt{g}\{ f(\phi) R [g]           \label{act}
                 + h(\phi)g\MN\phi_{,\mu}\phi_{,\nu} - F\MN F\mn\}.
\nnn
\ear
   Here $R[g]$ is the scalar curvature, $g = |\det (g\mn)|$, $f$ and $h$
   are certain functions of $\phi$, varying from theory to theory. Exact
   \ssph\ solutions for this system are well known \cite{penney,br73}, but
   their qualitative behaviour is rather diverse and depends on the nature
   of the functions $f$ and $h$.

   Wormholes form one of the generic classes of solutions in theories where
   the kinetic term in (\ref{act}) is negative \cite{br73} (more precisely,
   if $l(\phi)$, defined in (\ref{ps-f}), is negative). A particular case
   of this kind of wormholes, namely, wormholes with a ``ghost''
   massless minimally coupled scalar field in \GR\ [\eq(\ref{act}),
   $f(\phi)\equiv 1,\ h(\phi)\equiv -1$] was considered by H. Ellis
   \cite{h_ellis}.

   The energy conditions, NEC in particular, are, however, violated as well
   by ``less exotic'' sources, such as the so-called nonminimally coupled
   scalar fields in \GR, represented by the action (\ref{act}) with the
   functions
\beq                                                        \label{nonmin}
      f(\phi) = 1-\xi \phi^2, \quad \xi =\const; \qquad h(\phi) \equiv 1.
\eeq

   Scalar-vacuum (with $F\mn=0$) \ssph\ \wh\ solutions were found in such a
   theory in \Ref{br73} (and were recently discussed in \Ref{bar-vis99})
   for conformal coupling, $\xi=1/6$, and in \Ref{bar-vis00} for any $\xi >
   0$. The easiness of violating the energy conditions, so evident due to
   the appearance of \wh\ solutions, even made Barcelo and Visser discuss a
   ``restricted domain of application of the energy conditions''
   \cite{bar-vis00}. We recently proved \cite{bg01} that
   all these scalar-vacuum wormhole solutions are unstable under \sph\
   perturbations. The instability turns out to be of catastrophic nature:
   the increment of perturbation growth has no upper bound, hence, within
   linear perturbation theory, such a \wh, if once formed, should decay
   immediately and instantaneously. A full dynamical solution (yet to be
   found) would probably show a finite but still enormous decay rate.

   The purpose of this paper is to extend these results to charged \whs. We
   will show in \sect 2 that among the electrovacuum \ssph\ solutions of
   the theory (\ref{act}), (\ref{nonmin}) there is, for any $\xi>0$, a
   4-parameter family of \wh\ solutions. (For $\xi = 1/6$ this is already
   known from \cite{br73}.) The parameters can be identified as the mass,
   the electric and magnetic charges and the scalar field value at infinity.
   One more parameter, the scalar charge, is expressed in terms of the
   others. The instability of these \whs\ is demonstrated in \sect 3.

   As a tool, we use a transition to the Einstein conformal frame, in which
   the scalar field is minimally coupled to gravity. In all the \wh\
   solutions, the full manifold $\MJ[g]$ where the theory (\ref{act}) is
   formulated, maps to two Einstein-frame manifolds separated by the sphere
   \Str\ where $f=0$, and the instability develops in the neighbourhood
   of this sphere.

\section{Charged wormhole solutions}

\subsection {The general static solution}

   The general STT action (\ref{act}) is
   simplified by the well-known conformal mapping \cite{Wagoner1970}
\beq
   g\mn = \og\mn/|f(\phi)|,                             \label{trans-g}
\eeq
   accompanied by the scalar field transformation $\phi\mapsto \psi$ such
   that
\beq                                                       \label{ps-f}
   \frac{d\psi}{d\phi}= \pm \frac{\sqrt{|l(\phi)|}}{f(\phi)},
      \qquad    l(\phi) \eqdef fh +\frac 32
                        \biggl(\frac{df}{d\phi}\biggr)^2.
\eeq
   In terms of $\og\mn$ and $\psi$ the action takes the form
\bearr                                                   \label{act-E}
    S = \int d^4 x \sqrt{|\og|} \Bigl\{ (\sign f) \Bigl[  R [\og]
\nnn\cm
    + \og \MN \psi_{,\mu} \psi_{\nu} \,\sign l(\phi)\Bigr]
                        - F^{\mu\nu}F_{\mu\nu}   \Bigr\}
\ear
   (up to a boundary term which does not affect the field
   equations). Here $R[\og]$ is the Ricci scalar obtained from $\og\mn$, and
   the indices are raised and lowered using $\og\mn$. The electromagnetic
   field Lagrangian is conformally invariant, and $F\mn$ is not transformed.

   The space-time $\MJ[g]$ with the metric $g\mn$ is referred to as the
   {\it Jordan conformal frame}, generally regarded to be the physical
   frame in STT; the {\it Einstein conformal frame\/} $\ME[\og]$ with the
   field $\psi$ then plays an auxiliary role. The action (\ref{act-E})
   corresponds to conventional \GR\ if $f>0$, and the normal sign of scalar
   kinetic energy is obtained for $l(\phi) > 0$.

   The general \ssph\ solution to the Einstein-Maxwell-scalar equations that
   follow from (\ref{act-E}), was first found by Penney \cite{penney} and in
   a more complete form in \cite{zay,brpr}. Let us write it in the form
   suggested in \cite{br73}, restricting ourselves to the ``normal'' case
   $f >0$, $l>0$:
\bear
    ds_{\rm E}^2  \eql \e^{2\gamma(u)} dt^2
        -\e^{2\alpha(u)}du^2 - \e^{2\beta(u)} d\Omega^2       \label{ds-E}
\nnn \nqq
   = \frac {q^{-2}dt^2}{s^2(h, u+u_1)}
   - \frac{q^2 s^2(h,u+u_1)}{s^2(k,u)}
    \biggl[\frac{du^2}{s^2(k,u)}+d\Omega^2\biggr],
\nnn \\
     \psi(u)\eql Cu + \psi_1,                              \label{psi}
\\
    F_{01} \eql - F_{10} =
       q \,\e^{\alpha + \gamma - 2\beta}                   \label{F}
\\
      \eql [q\, s^2(h, u+u_1)]^{-1},                       \label{F01}
\ear
   where the subscript ``E'' stands for the Einstein frame;
   $d\Omega^2= d\theta^2 + \sin^2\theta d\varphi^2$ is the linear element
   on a unit sphere; $q=q_e$ (the electric charge), $C$ (the scalar charge),
   $h$, $k$ and $\psi_1$ are real integration constants.  The function
   $s(k,u)$ is defined as follows:
\beq                                                     \label{s}
    s(k,u) = \vars     {
                    k^{-1}\sinh ku,  \ & k > 0 \\
                                 u,  \ & k = 0 \\
                    k^{-1}\sin ku,   \ & k < 0.  }
\eeq

   Here $u$ is a convenient radial variable (it is a harmonic coordinate
   in the Einstein frame, $\DAL u =0$). The range of $u$ is $0 < u < \umx$,
   where $u=0$ corresponds to spatial infinity, while $\umx$ may be finite
   or infinite depending on the constants $k$, $h$ and $u_1$.

   The integration constants are related by
\bear                                                       \label{khC}
    2k^2 \sign k \eql 2h^2 \sign h + C^2,
\\
    s^2(h,\ u_1) \eql 1/q^2.                                  \label{u1}
\ear
   The latter condition, preserving some discrete arbitrariness of $u_1$,
   provides the natural choice of the time scale ($\og_{00}=1$) at spatial
   infinity ($u=0$). Without loss of generality we put $C > 0$ and
   $\psi_1=0$.

   As usual, in addition to the electric field $F_{01}=-F_{10}$
   given by (\ref{F01}), one can include a radial magnetic field
   $F_{32}=-F_{23} = q_m \sin\theta$ where $q_m$ is the magnetic charge.
   One should then understand $q^2$ in (\ref{ds-E}), (\ref{u1}) and
   further on as $q^2 = q_e^2 + q_m^2$, in (\ref{F}) one should replace
   $q$ with $q_e$ and in (\ref{F01}) $1/q$ with $q_e/q^2$. In what follows,
   we will bear in mind this opportunity without special mentioning.

   The solution contains four essential integration constants:
   $k$ or $h$ and the charges $q_e,\ q_m$ and $C$. The mass $M$
   in the Einstein frame is obtained by comparing the asymptotic of
   (\ref{ds-E}) at small $u$ with the Schwarzschild metric:
\beq
     GM = \pm \sqrt{q^2 + h^2 \sign h}                         \label{GM}
\eeq
   where $G$ is Newton's gravitational constant. The ``$\pm$'' sign
   depends on the choice of $u_1$ among the variants admitted by (\ref{u1}).

   The Reissner-Nordstrom solution of \GR\ is a special case obtained
   herefrom by putting $C=0$. Then from (\ref{khC}) it follows $h=k$, and the
   familiar form of the Reissner-Nordstrom metric is recovered after
   a transition to the curvature coordinates,
   $-\og_{\theta\theta} = r^2$:
\beq
    r = \frac{|q|\,s(k,u+u_1)}{s(k,u)}. 		      \label{RN}
\eeq

   To obtain another special case $q=0$ (the scalar-vacuum solution), one
   should consider the limit $q\to 0$ preserving the boundary condition
   (\ref{u1}). This is only possible for $k > h \geq 0$ and $u_1\to\infty$.
   The resulting metric is
\beq
   ds_{\rm E}^2 = \e^{-2hu} dt^2 - \frac{k^2 \e^{2hu}}{\sinh^2 (ku)}
      \biggr[\frac{du^2}{\sinh^2(ku)} + d\Omega^2 \biggl].   \label{g-vac}
\eeq
   The scalar field is determined, as before, from (\ref{psi}), and the
   integration constants are related by
\beq                                                      \label{k-vac}
            2k^2 = 2h^2 + C^2
\eeq
   It should be noted that in (\ref{g-vac}), (\ref{k-vac}) the constant $h$
   can have any sign, and for the mass $M$ we have simply $GM = h$.

   This is the Fisher solution \cite{fish} in terms of the harmonic $u$
   coordinate. Its more familiar form, used, in particular, in
   Refs.\,\cite{bar-vis99,bar-vis00}, corresponds to the coordinate $r$
   connected with $u$ by $r = 2k/(1-\e^{-2ku})$, and
   the metric in terms of $r$ has the form
\bearr
             ds^2_{\rm E}= (1-2k/r)^a dt^2              \label{g-vac'}
    \nnn \quad
         - (1-2k/r)^{-a}\bigl[dr^2 + r^2(1-2k/r)d\Omega^2\bigr],
\ear
   with $a=h/k$. The Schwarzschild solution is then recovered in case $C=0$,
   $a=1$.

   All the corresponding Jordan-frame solutions for $l(\phi)>0$ are obtained
   from (\ref{psi}), (\ref{ds-E}) using the transformation (\ref{trans-g}),
   (\ref{ps-f}).

\subsection {Continued solution in the Jordan frame}

   Let us now turn to wormhole solutions for the nonminimal coupling
   (\ref{nonmin}), $\xi > 0$. The transformation (\ref{ps-f}) takes the form
\beq                                                      \label{trans-f}
   \frac{d\psi}{d\phi}
            = \frac {\sqrt{|1-\phi^2(\xi-6\xi^2)|}}{1-\xi\phi^2},
\eeq
   where, without loss of generality, we have chosen the plus sign before
   the square root. We assume that spatial infinity in the Jordan space-time
   $\MJ$ corresponds to $|\phi| < 1/\sqrt{\xi}$, where $f(\phi) >0$, so that
   the gravitational coupling has its normal sign.

   Generically, the solution in $\ME[\og]$ has a naked singularity
   at $u=\umx$, and, though its nature can change
   due to the transformation to $g\mn$, it remains to be a singularity in
   $\MJ[g]$. An exception is the case when the solution is smoothly
   continued in $\MJ[g]$ through the sphere \Str\  ($u=\infty$,
   $\phi=1/\sqrt{\xi}$) which is singular in $\ME[\og]$ but regular in
   $\MJ[g]$. The infinity of the conformal factor $1/f$ thus compensates the
   zero of both $\og_{tt}$ and $\og_{\theta\theta}$ simultaneously.
   Wormhole solutions can only be found in this case. It
   happens when, in accord with (\ref{khC}),
\beq                                                        \label{k2h}
    k = 2h =2C/\sqrt{6} >0,\cm    u_1>0,
\eeq
   which selects a special subfamily among all solutions. We will restrict
   our attention to this subfamily. Note that now
   $s(k,u)= (2h)^{-1}\sinh (2hu)$, \
   $s(h, u+u_1) = h^{-1}\sinh (hu+hu_1)$ and $\umx=\infty$. According to
   (\ref{psi}) and (\ref{trans-f}),  we have $\psi\to\infty$ as $\phi\to
   1/\sqrt{\xi}-0$.

   Under the condition (\ref{k2h}) the solutions with and without charge in
   \ME\  are conveniently written in isotropic coordinates. Indeed, putting
   $y = \tanh (hu)$, we obtain:
\bear             \nq \label{dsEy}
     ds_{\rm E}^2 \eql \frac{(1-y^2) y_1^2}{(y+y_1)^2} \biggl[ dt^2
        -h^2\frac{(y+y_1)^4}{y_1^4\, y^4}(dy^2 + y^2 d\Omega^2)\biggr],
\nnn
\\
    \psi \eql \frac{\sqrt{6}}{2} \ln \frac{1+y}{1-y},      \label{psi-y}
\\                                                          \label{F-y}
    F_{01} \eql - F_{10} = \frac{q_e}{h}\,\frac{y_1^2}{(y+y_1)^2},
\ear
   where
\beq
    y_1 = \tanh (hu_1) = \frac{h}{\sqrt{h^2 + q^2}}.       \label{y1}
\eeq
   The vacuum solution is included here as the special case $q=0$, $y_1 =1$.
   The range of $u$, $u\in \R_+$, is converted into $y\in (0,1)$ where $y=0$
   corresponds to spatial infinity and $y=1-0$ to a naked singularity.

   To proceed to the Jordan frame, let us integrate
   \eq(\ref{trans-f}). This gives \cite{bar-vis00}%
\footnote
        {We have changed the notations as compared with
        \cite{bar-vis00}, in particular, we have replaced
        $\Phi_{\xi} \mapsto \sqrt{6}\phi$, $H\mapsto 1/H$ and $F^2\mapsto
	1/B$, to avoid imaginary $F$ at $\phi>1/\sqrt{\xi}$.}
\beq                                                          \label{ln}
        \psi = - \sqrt{3/2}\ln [B(\phi)H^2 (\phi)]
\eeq
    where
\beq                                                            \label{B}
   B(\phi) = B_0
        \frac{\sqrt{1-\eta\phi^2} - \sqrt{6}\xi\phi}
             {\sqrt{1-\eta\phi^2} + \sqrt{6}\xi\phi},
\eeq
    $B_0=\const$, while $H(\phi)$ is different for different $\xi$:
\bearr \nq                                                       \label{H}
     0<\xi<1/6:
\nnn
     H(\phi) = \exp\left[-\frac{\sqrt{1-6\xi}}{\sqrt{6\xi}}
                \arcsin \sqrt{\eta}\phi\right],
\nnn  \nq
     \xi > 1/6:
\nnn
     H(\phi) = \left[\sqrt{-\eta}\,\phi
                    +\sqrt{1-\eta\phi^2}\right]
            		^{\fract{\sqrt{6\xi-1}}{\sqrt{6\xi}}},
\ear
    where $\eta=\xi(1-6\xi)$, and $H\equiv 1$ for $\xi=1/6$.
    The function $H(\phi)$ is finite in the whole range of $\phi$ under
    consideration.

    \eq (\ref{ln}) is valid for $\phi<1/\sqrt{\xi}$, and the Jordan-frame
    metric $g\mn = \og\mn/f$ under the condition (\ref{k2h}) can be written
    in terms of the coordinate $y$ as follows:
\bearr 		                                                \label{ds-J}
     ds^2_{\rm J} =
               \frac{BH^{2}}{1-\xi\phi^2}
    			\biggl[ \frac{(1+y)^2}{(y+y_1)^2}y_1^2dt^2
\nnn \cm
        -h^2\frac{(1+y)^2 (y+y_1)^2}{y_1^2\, y^4}
				(dy^2 + y^2 d\Omega^2)\biggr],
\ear
    where $y$ can be expressed in terms of $\phi$:
\beq
	y = \frac{1-BH^2}{1+BH^2}.                              \label{yBH}
\eeq

    The metric is thus actually expressed in terms of the scalar field
    $\phi$ used as a coordinate. The isotropic coordinate $y$ conveniently
    shortens the expression (\ref{ds-J}) and makes it easy to see that the
    metric, originally built for $\phi < \phio$ ($y<1$), is smoothly
    continued across the surface \Str\ ($\phi=\phio,\ y=1$). Indeed, in a
    close neighbourhood of \Str, for $\phi = (\phi - \delta)/\sqrt{\xi}$
    with $\delta \ll 1$ one has
\[
     B \approx B_0 \delta/(12\xi),\qquad
     1 -\xi\phi^2 \approx 2\delta\,
\]
    whence
\beq                                                         \label{factor}
     \frac{BH^2}{1-\xi\phi^2}\ \bigg|_{y=1} =
			\frac{B_0}{24\xi} H^2\bigg|_{\phi=\phio}.
\eeq
    It is easily shown that this ratio is not only finite on \Str\ but also
    smoothly changes across it, so that \eq (\ref{ds-J}) comprises an
    analytic continuation of the metric, obtained from
    (\ref{ds-E})--(\ref{F01}) in case (\ref{k2h}) by the transformation
    (\ref{trans-g}), (\ref{ps-f}), beyond \Str. The coordinate
    $y$ covers the whole manifold $\MJ[g]$, and it is now possible to study
    the properties of the system as a whole.

    Before doing that, let us note that the new region $\phi > \phio$
    ($y>1$) in \MJ\ can also be obtained by the same transformation
    (\ref{trans-g}), (\ref{ps-f}) from a certain Einstein frame.  An
    essential difference from the previous solution is that, since
    $f(\phi)$ is now negative, (\ref{act-E}) leads to the Einstein equations
    with a reversed sign of the electromagnetic energy-momentum tensor. As a
    result, the solution in this second Einstein-frame manifold%
\footnote
       {The prime will designate quantities describing the Einstein frame
         or $\phi > 1/\sqrt{\xi}$.  }
    $\ME'$ will have the same form (\ref{ds-E})--(\ref{F01}), but with the
    replacement
\beq
	s(h,u+u_1)\ \mapsto \ h'{}^{-1} \cosh (h'u + h'u_1),    \label{E'1}
\eeq
    where $h'>0$, and the relation (\ref{khC}) is replaced by $2k'{}^2 =
    2h'{}^2 + C'{}^2$ where $k' >0$.

    The solution in $\ME'$ is also regularized by the factor $1/f$ on
    \Str, and the integration constants in it satisfy the condition
    $k'=2h'$, similar to (\ref{k2h}). Other integration constants are
    adjusted as well, in particular, the charges $q_e$ and $q_m$ are the
    same on both sides of \Str, providing the continuity of the
    electromagnetic field.

\subsection{Wormhole solutions}

    Let us begin with the simplest case $\xi=1/6$ (conformal coupling).
    Then instead of (\ref{ln})--(\ref{H}) one can write for
    $\phi< \sqrt{6}$
\beq
    \phi = \sqrt{6}\tanh [(\psi + \psi_0)/\sqrt{6}],
            \cm \psi_0 = \const,               \label{phi-1/6}
\eeq
    where $\psi=Cu$ and due to (\ref{k2h}) $C= h\sqrt{6}$.
    The Jordan-frame solution in terms of the isotropic
    coordinate $y$ takes the form \cite{br73}
\bear
     ds_{\rm J}^2 \eql \frac{(1+yy_0)^2}{1-y_0^2}           \label{ds6}
      \biggl [\frac{y_1^2\, dt^2}{(y+y_1)^2}
\nnn \cm\
    -h^2\,\frac{(y+y_1)^2}{y_1^2 y^4}(dy^2 + y^2 d\Omega^2)\biggr],
\\
     \phi \eql \sqrt{6}\, \frac{y+y_0}{1 + yy_0},           \label{phi6}
\ear
    where $y_0 = \tanh (\psi_0/\sqrt{6})$ and $y_1 \in (0,1)$;
    the expressions for $F\mn$ are evident.

    The original Einstein-frame solution corresponds to $y<1$, $y=0$ is
    spatial infinity while the sphere $y=1$ is \Str, where the solution
    (\ref{ds6}), (\ref{phi6}) is manifestly regular. The region $y>1$ is an
    analytic continuation of the solution in $\MJ[g]$ to $\phi > \sqrt{6}$
    and corresponds to another Einstein-frame solution described above.

    The properties of the solution at $y>1$ depend on the constant $y_0$
    which characterizes the $\phi$ field at spatial infinity. Namely, if
    $y_0 < 0$, then the solution has a naked singularity at $y = -1/y_0 >1$.
    If $y_0=0$, we obtain a black hole with electromagnetic and scalar
    charges \cite{brpr,br73,bek74}; introducing $r= h(y+y_1)/(y_1 y)$, we
    obtain
\bear
     ds^2 \eql (1-m/r)^2 dt^2 - (1-m/r)^{-2}dr^2 -r^2 d\Omega^2,   \label{BH}
\nn
     \phi \eql C/(r-m)
\ear
    where $m= GM = \sqrt{h^2 + q^2},\ C = \sqrt{6}h$.
    On the horizon, $r=m$, despite $\phi\to\infty$, the energy-momentum
    tensor of the scalar field is finite. This solution (mainly its
    neutral special case $q=0$) was repeatedly discussed as an interesting
    counterexample of the well-known no-hair theorems; its instability
    under \sph\ perturbations has been proved in \Ref{br78}.

    Lastly, if $y_0 >0$, then $y$ ranges from 0 to
    $\infty$, and $y=\infty$ is another flat spatial infinity. This is the
    sought-for \wh\ solution, parametrized by the four constants $h$, $q_e$,
    $q_m$ and $y_0$. The position and radius of the \wh\ neck (minimum of
    $r^2 = -g_{\theta\theta}$) are given by
\beq                                                            \label{neck}
      y_{\rm neck} = \frac{\sqrt{y_1}}{\sqrt{y_0}}, \cm
      r_{\rm neck} =
         \frac{h (1 + \sqrt{y_0 y_1})^2}{y_1\sqrt{1-y_0^2}} .
\eeq

    For $\xi \ne 1/6$ the analytical relations are much more complicated,
    but the qualitative behaviour of the solution can be described
    rather easily.

    In case $\xi > 1/6$, for any $B_0$, with growing $\phi$ the quantity
    $B^2 H^{-4}$ eventually reaches the value 1, where $g_{\theta\theta}\to
    \infty$, i.e., we arrive at another spatial asymptotic, and it is
    straightforward to verify that this infinity is flat. In other words, we
    obtain again a static \wh.

    In case $\xi < 1/6$ everything depends on $B_0$. If
\beq                                                           \label{B0}
    B_0 < B_0^{\rm cr}
           =\exp\biggl(-\pi\sqrt{\frac{1-6\xi}{6\xi}}\biggr),
\eeq
    the situation is the same as for $\xi> 1/6$, i.e., a \wh. If
    $B_0 > B_0^{\rm cr}$, then, while $g_{\theta\theta}$ is still finite,
    $\phi$ reaches the value $1/\sqrt{\eta} = 1/\sqrt{\xi(1-6\xi)}$, the
    location of a curvature singularity \cite{bar-vis00}. So we have a naked
    singularity instead of a \wh. Lastly, for $B_0 = B_0^{\rm cr}$, the
    maximum value of $\phi$ is again $1/\sqrt{\eta}$, but now it is
    non-flat spatial infinity.

\section{Stability analysis}

\subsection{Perturbation equations}

    Consider small (linear) spherically symmetric perturbations of the above
    \wh{}s. It is helpful to work separately in each of the two
    Einstein-frame manifolds $\ME$ and $\ME'$, perturbing the metric
    quantities $\alpha,\ \beta,\ \gamma$ in (\ref{ds-E}) and the field
    $\psi$, replacing
\beq                                                        \label{st1}
    \psi(u) \to \psi(u,t)=\psi(u) + \delta\psi(u,t)
\eeq
    and similarly for other quantities; the same is done for their
    counterparts in $\ME'$. Due to spherical symmetry, the only dynamical
    degree of freedom is the scalar field, obeying the equation $\DAL \psi
    =0$, while other perturbations must be expressed in terms of
    $\delta\psi$ and its derivatives via the Einstein equations. The
    perturbed scalar equation has the form
\beq                                                          \label{st2}
    \e^{-\gamma+\alpha+2\beta}\ddot{\psi}-
                      \left(\e^{\gamma-\alpha+2\beta}\psi_u\right)_u = 0.
\eeq
    where the dot stands for $\d/\d t$ and the subscript $u$ for the radial
    coordinate derivative $\d/\d x^1$. One can notice that \eq(\ref{st2})
    decouples from perturbations other than $\delta\psi$ if one chooses the
    frame of reference and the coordinates in the perturbed space-time
    (the gauge for short) so that
\beq
    \delta\alpha = 2\delta\beta + \delta\gamma.           \label{st3}
\eeq
    The relation $\alpha=2\beta + \gamma$ thus holds for both the static
    background written as in (\ref{psi}), (\ref{ds-E}) and the
    perturbations. The unperturbed part of \eq (\ref{st2}) reads
    $\psi_{uu}=0$ and is satisfied by (\ref{psi}), while for $\delta\psi$ we
    obtain the wave equation
\beq
    \e^{4\beta(u)} (\delta\psi)\,\ddot{} - \delta\psi_{uu}=0.    \label{st4}
\eeq
    The static nature of the background solution makes it possible to
    separate the variables,
\beq
    \delta\psi = \Phi(u) \e^{i\omega t},                      \label{Phi}
\eeq
    and to reduce the stability problem to a boundary-value problem for
    $\Phi(u)$. Namely, if there exists a nontrivial solution to
    (\ref{st4}) with $\omega^2 <0$, satisfying some physically
    reasonable boundary conditions, then the static background system is
    unstable since the perturbations can exponentially grow with $t$.
    Otherwise it is stable in the linear approximation.

    Suppose $-\omega^2 = \Omega^2,\ \Omega > 0$. The equation that follows
    directly from (\ref{st4}),
\beq
    \Phi_{uu} - \Omega^2 \e^{4\beta(u)}\Phi=0,              \label{ePhi}
\eeq
    is converted to the normal Liouville (Schr\"odinger-like) form
\bearr
    d^2 Y/dx^2 - [\Omega^2+V(x)] Y(x) =0,   \nnn \cm
    V(x) = \e^{-4\beta}(\beta_{uu}-\beta_u{}^2).            \label{schrod}
\ear
    by the transformation
\beq
    \Phi(u) = Y(x)\e^{-\beta},\qquad                        \label{u2x}
                            x = - \int \e^{2\beta(u)}du.
\eeq

    \eq (\ref{schrod}) makes it possible to use the experience of
    quantum mechanics (QM): $\Omega^2$ here corresponds to $ - E$ in the
    Schr\"{o}dinger equation. In other words, the presence of ``negative
    energy levels'' $E= - \Omega^2 <0$ for the potential $V(x)$ indicates
    the instability of our system.

    The variable $x$ behaves as follows at small and large $u$:
\begin{description}
\item [] $u\to 0$ (spatial infinity): $x\approx \e^\beta\approx 1/u$;
\item [] $u\to \infty$ (the sphere \Str): $x\approx 8h \e^{-2hu}$.
\end{description}
    For the potential $V(x)$ one finds:
\bearr                                                        \label{asV}
     V(x) \approx 2h/x^3 \quad
            (x\to\infty \quad \mbox{--- spatial asymptotic}),
\nnn
     V(x) \approx -1/(4x^2) \
                (x\to 0 \ \mbox{--- the sphere \Str.}).
\ear
    Thus we have a quadratic potential well at \Str, which is placed at $x=0$
    by choosing the proper value of the arbitrary constant in the definition
    of $x$ in \eq (\ref{u2x}).

    The same form of \eq (\ref{schrod}) is obtained for the Einstein frame
    $\ME'$, but with another potential $V(\phi)$ due to the slightly
    different form of the solution in this ``antigravitational'' region.
    One easily finds, however, that the asymptotics of the potential
    at $x\to 0$ and $x\to\infty$ are again given by (\ref{asV}), though
    with $h$ replaced by some $h'>0$ which is, in general, not equal to $h$.

    It makes sense to change $x\to -x$ in $\ME'$, which does not affect \eq
    (\ref{schrod}) but makes it possible to unify the perturbation equations
    for the two parts of $\MJ$, the space-time of the Jordan-frame. We thus
    obtain \eq (\ref{schrod}) with $x \in \R$ and a certain function
    $V(x)$, vanishing at large $|x|$ and providing a potential well of the
    form $V\approx 1/(4x^2)$ near $x=0$.

    The boundary conditions at both spatial asymptotics are obtained from
    the requirement that the perturbations should possess finite energy.
    This requirement upon the perturbed EMT leads to the condition
    $xY\to 0$ as $x\to\pm\infty$. Meanwhile, the
    asymptotic form of any solution of (\ref{schrod}) with $\Omega >0$
    at large $|x|$ is
\beq                                                          \label{asymp}
    Y\approx C_1\e^{\Omega |x|}+C_2\e^{-\Omega |x|}, \qquad
    C_{1,2} = \const.
\eeq
    Therefore an admissible solution is the one with $C_1=0$, with
    only a decaying exponential. Actually, the conditions at both
    infinities are that $Y\to 0$, i.e., coincide with the boundary
    conditions for the one-dimensional wave function under the
    same potential in QM.

    As is evident from QM (see, e.g., \cite{Rybakov}), a potential well of
    the form $V\approx 1/(4x^2)$ always possesses negative energy levels,
    $E = -\Omega^2 < 0$; moreover, the absolute value of $\Omega$ has no
    upper bound. The latter statement can be proved, e.g., by comparing
    \eq (\ref{schrod}) with our $V(x)$ and with rectangular potentials
    $\tilde V \geq V$ for which $Y(x)$ and $E$ are easily found; one can then
    use the fact that $E_{\min}[V] < E_{\min}[\tilde V]$ where $E_{\min}$ is
    the lowest energy level (ground state) for a given potential.

    Recalling that $\Omega$ is the perturbation growth increment, we can
    conclude that our \wh{}s decay instantaneously within linear
    perturbation theory. Nonperturbative analysis would probably smooth out
    this infinite decay rate.

    The behaviour of the perturbations near \Str\ ($x=0$) is of
    interest. The asymptotic form of the solution to (\ref{schrod}) at small
    $x$ is
\beq
	Y = \sqrt{|x|} (c_1 + c_2 \ln |x|), \quad c_{1,2} =\const,
\eeq
    therefore the perturbation $\delta\psi \sim \Phi \sim Y/\sqrt{|x|}$
    behaves as $c_1 + c_2 \ln |x|$, i.e., generically blows up at $x=0$ but
    at the same rate as $\psi$ itself, so that the perturbation scheme
    still works. Furthermore, with (\ref{trans-f}) it is easy to find that
    the perturbation $\delta\phi$ behaves at small $x$ as
    $x(c_1 + c_2 \ln |x|)$, so that $\delta\phi(0) =0$.
    In other words, the perturbation as a function of time rapidly grows
    around \Str\ due to the instability but vanishes on this sphere itself.

\section{Concluding remarks}

    Our perturbation analysis proves the instability of both neutral and
    charged \wh\ solutions for any $\xi > 0$. This conclusion can be
    extended to all solutions continued through the sphere \Str, where the
    original function $f(\phi)$ in the action (\ref{act}) vanishes.
    This actually means that the effective gravitational constant,
    proportional to $f^{-1}$, blows up and changes its sign.  The violent
    instability occurs due to a negative pole of the perturbational
    effective potential $V(x)$ on \Str. Comparing the present results with
    those of our previous paper \cite{bg01}, we see that neither the
    particular value of the coupling constant $\xi$ nor the presence of
    matter (the electromagnetic field in our case) change the situation.

    The instability of black holes with a conformal scalar field, found
    long ago in \Ref {br78}, is another example of such a phenomenon.
    A similar instability was pointed out by Starobinsky \cite{star81} for
    cosmological models with conformally coupled scalar fields.

    It is quite plausible that instabilities of this kind are a common
    feature of STT solutions with conformal continuations, for which the
    transformation (\ref{trans-g}) maps the Einstein-frame manifold
    $\ME[\og]$ to only a part of the whole Jordan-frame manifold $\MJ[g]$.
    Solutions containing such continuations always exist in STT in which
    the function $f(\phi)$ has at least one simple zero \cite{vac4},
    irresespective of the particular form of $f(\phi)$. Wormholes solutions
    turn out to be generic among the conformally continued solutions, but
    there also exist other kinds of configurations \cite{vac4}. One can
    anticipate that all of them are unstable since the cause of instability
    is the transition itself rather than the wormhole nature of the
    solutions.

\small
\Acknow{KB is thankful to Prof. Mario Novello for kind
    hospitality at CBPF and numerous helpful discussions. We acknowledge
    partial financial support from the Ministry of Industry, Science and
    Technology of Russia and the Ministry of Education of Russia.}

\end{document}